\documentclass[twocolumn,twocolappendix]{aastex7}

\usepackage{CJKutf8}
\usepackage{url}

\makeatletter
\DeclareRobustCommand{\HI}{%
  \mbox{H\check@mathfonts\fontsize\sf@size\z@\selectfont I}%
}
\makeatother
\newcommand{\lya}{Ly$\alpha$}
\newcommand{\lyb}{Ly$\beta$}

\shorttitle{Higher $\xi_{\rm ion}$ in Overdense Regions}
\shortauthors{Zhu et al.}

\begin{document}

\title{SMILES: Potentially Higher Ionizing Photon Production Efficiency in Overdense Regions}

\author[0000-0003-3307-7525]{Yongda Zhu}
\affiliation{Steward Observatory, University of Arizona, 933 North Cherry Avenue, Tucson, AZ 85721, USA}
\email{yongdaz@arizona.edu}

\author[0000-0002-8909-8782]{Stacey Alberts}
\affiliation{Steward Observatory, University of Arizona,
933 North Cherry Avenue, Tucson, AZ 85721, USA}
\email{}

\author[0000-0002-6221-1829]{Jianwei Lyu}
\affiliation{Steward Observatory, University of Arizona,
933 North Cherry Avenue, Tucson, AZ 85721, USA}
\email{}

\author[0000-0002-9288-9235]{Jane Morrison}
\affiliation{Steward Observatory, University of Arizona, 933 North Cherry Avenue, Tucson, AZ 85721, USA}
\email{}

\author[0000-0003-2303-6519]{George H.\ Rieke}
\affiliation{Steward Observatory, University of Arizona, 933 North Cherry Avenue, Tucson, AZ 85721, USA}
\email{}

\author[0000-0001-6561-9443]{Yang Sun}
\affiliation{Steward Observatory, University of Arizona, 933 North Cherry Avenue, Tucson, AZ 85721, USA}
\email{}

\author[0000-0003-4337-6211]{Jakob M.\ Helton}
\affiliation{Steward Observatory, University of Arizona, 933 North Cherry Avenue, Tucson, AZ 85721, USA}
\email{}

\author[0000-0001-7673-2257]{Zhiyuan Ji}
\affiliation{Steward Observatory, University of Arizona, 933 North Cherry Avenue, Tucson, AZ 85721, USA}
\email{}
 
\author[0000-0003-0883-2226]{Rachana Bhatawdekar}
\affiliation{
European Space Agency (ESA), European Space Astronomy Centre (ESAC), Camino Bajo del Castillo s/n, 28692 Villanueva de la Cañada, Madrid, Spain}
\email{}

\author[0000-0001-8470-7094]{Nina Bonaventura}
\affiliation{Steward Observatory, University of Arizona, 933 North Cherry Avenue, Tucson, AZ 85721, USA}
\email{}

\author[0000-0002-8651-9879]{Andrew J.\ Bunker }
\affiliation{Department of Physics, University of Oxford, Denys Wilkinson Building, Keble Road, Oxford OX1 3RH, UK}
\email{}

\author[0000-0001-6052-4234]{Xiaojing Lin}
\affiliation{Department of Astronomy, Tsinghua University, Beijing 100084, China}
\affiliation{Steward Observatory, University of Arizona, 933 North Cherry Avenue, Tucson, AZ 85721, USA}
\email{}

\author[0000-0002-7893-6170]{Marcia J. Rieke}
\affiliation{Steward Observatory, University of Arizona, 933 North Cherry Avenue, Tucson, AZ 85721, USA}
\email{}

\author[0000-0002-5104-8245]{Pierluigi Rinaldi}
\affiliation{Steward Observatory, University of Arizona, 933 North Cherry Avenue, Tucson, AZ 85721, USA}
\email{}

\author[0000-0003-4702-7561]{Irene Shivaei} \affiliation{Centro de Astrobiolog\'ia (CAB), CSIC-INTA, Ctra. de Ajalvir km 4, Torrej\'on de Ardoz, E-28850, Madrid, Spain}
\email{}

\author[0000-0001-9262-9997]{Christopher N.\ A.\ Willmer}
\affiliation{Steward Observatory, University of Arizona, 933 North Cherry Avenue, Tucson, AZ 85721, USA}
\email{}

\author[0000-0002-1574-2045]{Junyu Zhang}
\affiliation{Steward Observatory, University of Arizona, 933 North Cherry Avenue, Tucson, AZ 85721, USA}
\email{}

\correspondingauthor{Yongda Zhu}
\email{yongdaz@arizona.edu}

\begin{abstract}
The topology of reionization and the environments where galaxies efficiently produce ionizing photons are key open questions. For the first time, we investigate the \added{trend} between ionizing photon production efficiency, $\xi_{\rm ion}$, and galaxy overdensity, $\log(1+\delta)$. We analyze the ionizing properties of \added{79} galaxies between \added{$1.0 < z < 5.2$} using JWST NIRSpec medium-resolution spectra from the Systematic Mid-infrared Instrument (MIRI) Legacy Extragalactic Survey (SMILES) program. Among these, 67 galaxies have H$\alpha$ coverage, spanning \added{$1.0 < z < 3.1$}. The galaxy overdensity, $\log(1+\delta)$, is measured using the JADES photometric catalog, which covers the SMILES footprint. For the subset with H$\alpha$ coverage, we find that $\log\xi_{\rm ion}$ is positively correlated with $\log(1+\delta)$, with a slope of $0.94_{-0.46}^{+0.46}$. Additionally, the mean $\xi_{\rm ion}$ for galaxies in overdense regions ($\log(1+\delta) > 0.1$) is 2.43 times that of galaxies in lower density regions ($\log(1+\delta) < 0.1$). This strong \added{trend} is found to be independent of redshift evolution. Furthermore, our results confirm the robust correlations between $\xi_{\rm ion}$ and the rest-frame equivalent widths of the [O III] or H$\alpha$ emission lines. Our results suggest that galaxies in high-density regions are efficient producers of ionizing photons.
\end{abstract}

\section{Introduction} \label{sec:intro}
When and how reionization occurred carries important implications for the formation and evolution of the first stars, galaxies, and supermassive black holes. Recent observations from both quasar absorption lines and galaxies have made significant progress in constraining the timeline and topology of reionization \citep[see, e.g.,][for a review]{fan_quasars_2023}. Nevertheless, it remains challenging to build a self-consistent picture of cosmic reionization that connects the emission of ionizing photons with their absorption by the intergalactic medium (IGM; e.g., \citealp{robertson_galaxy_2022}).

Firstly, regarding the timeline of reionization, cosmic microwave background (CMB) observations indicate a midpoint of reionization at $z \sim 7-8$ \citep[][see also \citealp{de_belsunce_inference_2021}]{planck_collaboration_planck_2020}. This midpoint is supported by, or consistent with, multiple observations. They include (1) the \lya\ damping wing in $z\gtrsim7$ quasar spectra \citep[e.g.,][]{banados_800-million-solar-mass_2018,davies_quantitative_2018,wang_significantly_2020,yang_poniuaena_2020,greig_igm_2021,greig_igm_2024}, (2) the decline in observed \lya\ emission from galaxies at $z>6$ \citep[e.g.,][and references therein, but see \citealp{jung_texas_2020,wold_lager_2021}]{stark_keck_2010,caruana_spectroscopy_2014,mason_universe_2018,mason_inferences_2019,hoag_constraining_2019,hu_ly_2019,jones_jades_2024,jones_jades_2024-1}, and (3) measurements of the thermal state of the IGM at $z>5$ \citep[e.g.,][]{boera_revealing_2019,gaikwad_consistent_2021}. Moreover, observations from quasar absorption lines suggest that reionization is patchy and may have ended as late as $z\sim5.3$, allowing galaxies sufficient time to produce the ionizing photons needed to complete the latter half of reionization (compared to an early end at $z=6$). These observations include (1) large-scale fluctuations in the \lya\ effective optical depth measured in quasar spectra \citep[e.g.,][]{fan_constraining_2006,becker_evidence_2015,eilers_opacity_2018,bosman_new_2018,bosman_hydrogen_2022,yang_measurements_2020}; (2) long troughs extending down to or below $z\simeq5.5$ in the \lya\ and \lyb\ forests \citep[e.g.,][]{becker_evidence_2015,zhu_chasing_2021,zhu_long_2022}, potentially indicating the existence of large neutral IGM islands \citep[e.g.,][]{kulkarni_large_2019,keating_long_2020,nasir_observing_2020,qin_reionization_2021}; (3) the evolution of metal-enriched absorbers at $z\sim6$ \citep[e.g.,][]{becker_evolution_2019,cooper_heavy_2019,davies_xqr-30_2023,davies_examining_2023,sebastian_e-xqr-30_2024}; (4) the dramatic evolution in the measured mean free path of ionizing photons over $5<z<6$ \citep[][see also \citealp{bosman_constraints_2021,gaikwad_measuring_2023,satyavolu_robustness_2023,roth_effect_2024,davies_constraints_2024}]{becker_mean_2021,zhu_probing_2023}; and (5) damping wing signals from stacked \lya\ forest at $z\sim 5.8$ \citep{zhu_damping_2024,spina_damping_2024}. Nevertheless, recent James Webb Space Telescope (JWST) observations of high-redshift galaxies may suggest very early starting and ending points of reionization, given the excessive ionizing photon budget \citep[see, e.g.,][]{munoz_reionization_2024,cain_chasing_2024}.

Equally puzzling is the correlation between IGM opacity and galaxy overdensity. Models and simulations aiming to explain the fluctuations in IGM opacity near the end of reionization require large-scale fluctuations in the metagalactic ionizing ultraviolet (UV) background \citep[e.g.,][but see \citealp{{daloisio_large_2015}}]{davies_large_2016,nasir_observing_2020}. In this scenario, highly opaque and transmissive IGM regions should be surrounded by galaxy under- and overdensities, respectively, since nearby galaxies are needed to boost the local UV background \citep[e.g.,][]{gangolli_correlation_2024,neyer_thesan_2024}. As expected, underdensities of galaxies traced by \lya-emitting galaxies are found around highly opaque quasar sightlines \citep[][]{becker_evidence_2018,christenson_constraints_2021,kashino_evidence_2020,ishimoto_physical_2022}. However, surprisingly, highly transmissive sightlines are also found to be associated with underdensities traced by Ly$\alpha$ emitters \citep[LAEs][]{christenson_relationship_2023}. Multiple studies have shown that LAEs avoid at least some of the highest density peaks \citep[e.g.,][]{kashikawa_habitat_2007,huang_evaluating_2022} or generally tend to trace lower-density regions \citep{cooke_nurturing_2013}. This discrepancy raises questions about whether late reionization models accurately capture the ionization of low-density regions or whether the ionizing photon budget depends on the galaxy environment.

One key component that could help resolve these puzzles is the ionizing photon production efficiency ($\xi_{\rm ion}$), which links the UV luminosity of galaxies to their contribution to reionization via the relation $\dot{N} = L_{\rm UV} \cdot \xi_{\rm ion} \cdot f_{\rm esc}$, where $\dot{N}$ is the rate of ionizing photon production, $L_{\rm UV}$ is the UV luminosity, and $f_{\rm esc}$ is the escape fraction of ionizing photons. Recent measurements of $\xi_{\rm ion}$ have shed light on its variation across different galaxy properties, such as the UV luminosity, equivalent width of H$\alpha$ and [O III] emission lines, etc. \citep[e.g.,][]{shivaei_mosdef_2018,tang_mmtmmirs_2019,prieto-lyon_production_2023,pahl_spectroscopic_2024,rinaldi_midis_2024,saxena_jades_2024,simmonds_ionizing_2023,simmonds_low-mass_2024,chevallard_physical_2018,castellano_ionizing_2023,begley_evolution_2025,harshan_canucs_2024,llerena_ionizing_2024}, but the relationship between $\xi_{\rm ion}$ and galaxy overdensity remains unexplored. This correlation is directly tied to the topology of reionization, as denser regions are expected to host more ionizing sources and could potentially reionize earlier or more efficiently than underdense regions.

The SMILES program \citep{alberts_smiles_2024,rieke_smiles_2024} offers a unique opportunity to investigate these questions at lower redshifts \added{(H$\alpha$ covering $0.7<z<3.7$)}, providing valuable insights into the epoch of reionization from a different cosmic time. As one of the largest spectral samples of cosmic-noon galaxies, SMILES focuses on the GOODS-S/HUDF \citep{giavalisco_great_2004, beckwith_hubble_2006} region and includes both MIRI and Near Infrared Spectrograph (NIRSpec) observations. The rich ancillary multi-wavelength coverage from X-ray to radio in this field allows for robust measurements of galaxy properties. Combined with high-quality Near Infrared Camera (NIRCam, \citealp{rieke_performance_2023}) photometry from JADES \citep{eisenstein_jades_2023,bunker_jades_2024,rieke_jades_2023,williams_jems_2023}, we can measure overdensities using photometric redshifts. This work presents the first measurement of the \added{trend} between $\xi_{\rm ion}$ and galaxy overdensity, with implications for our understanding of reionization.

This paper is structured as follows: Section \ref{sec:data} introduces the SMILES spectra used in this work; Section \ref{sec:methods} details the $\xi_{\rm ion}$ and overdensity measurements; Section \ref{sec:discussion} presents our results and discusses their implications; and Section \ref{sec:summary} summarizes our findings. We adopt the {\tt Planck18} cosmology \citep{planck_collaboration_planck_2020}, as implemented in {\tt astropy} \citep{astropy_collaboration_astropy_2022}, and all distances are quoted in comoving units.

\section{SMILES Spectra and Sample Selection}\label{sec:data}

\begin{deluxetable*}{cccccccccc}
\tablenum{1}
\tabletypesize{\footnotesize}
\tablecaption{Galaxy Spectra Used in this Work and Measured Properties}
\tablehead{
\colhead{ID} &
\colhead{RA} &
\colhead{DEC} &
\colhead{$z$} &
\colhead{$\log( 1 + \delta )$} &
\colhead{$M_{1500}$} &
\colhead{$\log({\rm H\alpha\ REW / \AA})$} &
\colhead{$\log({\rm [OIII]\ REW / \AA})$} &
\colhead{$O_{32}$} &
\colhead{$\log( \xi_{\rm ion, H\alpha} /\rm [Hz\,erg^{-1}] )$} \\\\
}
\decimalcolnumbers
\startdata
114544 & 53.18335 & -27.79602 & $4.949$ & $-0.03$ & $-18.82_{-0.28}^{+0.39}$ &       & $3.36_{-1.62}^{+0.30}$ & $6.26_{-1.44}^{+1.44}$ &       \\
1205   & 53.14898 & -27.78199 & $1.907$ & $0.10$  & $-22.61_{-0.28}^{+0.39}$ &       & $1.55_{-0.02}^{+0.02}$ & $0.49_{-0.03}^{+0.03}$ &       \\
1001   & 53.19236 & -27.79791 & $1.221$ & $0.31$  & $-19.65_{-0.28}^{+0.39}$ & $2.42_{-0.01}^{+0.01}$ & $1.96_{-0.01}^{+0.01}$ &        & $24.90_{-0.12}^{+0.16}$ \\
1095   & 53.18117 & -27.79099 & $2.740$ & $-0.00$ & $-20.83_{-0.28}^{+0.39}$ & $0.76_{-0.12}^{+0.09}$ &                       &        & $23.76_{-0.24}^{+0.25}$ \\
1108   & 53.19022 & -27.78993 & $2.481$ & $0.05$  & $-21.21_{-0.28}^{+0.39}$ & $2.35_{-0.01}^{+0.01}$ & $2.02_{-0.01}^{+0.01}$ & $1.21_{-0.03}^{+0.03}$ & $25.12_{-0.12}^{+0.16}$ \\
1118   & 53.17389 & -27.78857 & $1.674$ & $0.23$  & $-20.73_{-0.28}^{+0.39}$ & $2.24_{-0.01}^{+0.01}$ & $2.00_{-0.01}^{+0.01}$ &        & $24.91_{-0.12}^{+0.16}$ \\
1146   & 53.19760 & -27.78647 & $1.098$ & $0.25$  & $-22.78_{-0.28}^{+0.39}$ & $1.85_{-0.01}^{+0.01}$ & $0.85_{-0.07}^{+0.06}$ &        & $24.44_{-0.12}^{+0.16}$ \\
1152   & 53.19224 & -27.78608 & $2.454$ & $0.07$  & $-22.47_{-0.28}^{+0.39}$ & $2.30_{-0.01}^{+0.01}$ & $2.06_{-0.01}^{+0.01}$ & $1.09_{-0.03}^{+0.03}$ & $24.79_{-0.12}^{+0.16}$ \\
1155   & 53.17659 & -27.78552 & $1.317$ & $0.10$  & $-22.36_{-0.28}^{+0.39}$ & $1.42_{-0.01}^{+0.01}$ &                       &        & $24.47_{-0.13}^{+0.17}$ \\
1177   & 53.18605 & -27.78408 & $2.394$ & $-0.33$ & $-20.55_{-0.28}^{+0.39}$ & $2.37_{-0.01}^{+0.01}$ & $2.06_{-0.01}^{+0.01}$ & $1.18_{-0.03}^{+0.03}$ & $25.27_{-0.12}^{+0.16}$ \\
\enddata
\tablecomments{
Columns: 
(1) ID of the galaxy in SMILES program;
(2) \& (3) coordinates in J2000;
(4) photometric redshift;
(5) logarithm of the galaxy density contrast, $\log ( 1+ \delta) = \log\left( {n_{\rm g}}/{\langle n_{\rm g} \rangle} \right)$, where $n_{\rm g}$ is the local galaxy number density and $\langle n_{\rm g} \rangle$ is the mean number density;
(6) absolute UV magnitude at 1500 \AA\ after corrected for dust attenuation;
(7) logarithm of the rest equivalent width of the H$\alpha$ emission line;
(8) logarithm of the rest equivalent width of the [O III] emission line;
(9) oxygen line ratio, $O_{32} = \frac{[\rm OIII]\lambda5007}{[\rm OII]\lambda3727,3729}$;
(10) ionizing photon production efficiency derived from the H$\alpha$ emission line.
\\
(This table is available in its entirety in machine-readable form.) 
}
\label{tab:sample}
\end{deluxetable*}

The SMILES survey design is described in \citet{alberts_smiles_2024} and \citet{rieke_smiles_2024}. We used the eMPT tool \citep{bonaventura_near-infrared_2023} to configure the NIRSpec micro-shutter array (MSA) for optical spectroscopy follow-up of SMILES MIRI sources. A total of 168 unique targets were observed in the GOODS-S field. \added{All observed sources have MIRI coverage from the SMILES program, but not all were selected based on their MIRI properties.} The input catalog was assembled using multi-wavelength photometry and included diverse science priorities, such as star-forming galaxies, AGN, dusty systems, and quiescent galaxies. \added{Specifically, among the observed targets, 104 were flagged in the catalog as ``cosmic noon'' star-forming galaxies, 84 as AGN candidates \citep[see selections in][]{lyu_active_2024}, 24 as quiescent galaxies, and 86 as potential members of $z \gtrsim 5$ overdensities identified using FRESCO data \citep{helton_jwst_2023}. These categories are not mutually exclusive, and many targets belong to multiple science categories.} 

\begin{figure}[!ht]
    \centering
    \includegraphics[width=1\linewidth]{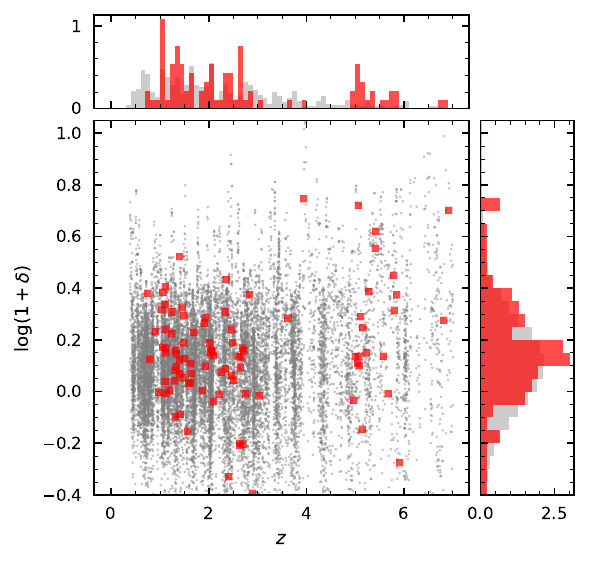}
    \caption{Overdensity and redshift distribution of galaxies used in this work (red data points and histograms). Spectra for galaxies at $z<3.1$ have the coverage of the H$\alpha$ line in our sample, while galaxies at high redshifts have [O III] covered. We also plot the overdensity and redshift distributions for JADES galaxies within the SMILES footprint in gray.
    }
    \label{fig:overdensity-z}
\end{figure}

\added{The full sample spans a wide redshift range ($0.3 < z < 7.3$) and UV luminosities from $M_{\rm UV} = -23$ to $-18$, with a median of $M_{\rm UV} \sim -20$. Sources were not selected based on known $\xi_{\rm ion}$ values or environmental information. This diverse and inclusive target list enables us to study ionizing photon production across a range of galaxy populations and environments. The redshift and overdensity (see Section \ref{sec:methods}) distributions of the final analysis sample are shown in Figure~\ref{fig:overdensity-z}.}

We used the G140M/F100LP and G235M/F170LP gratings to cover the wavelength range $0.97\text{--}3.07~\mu\text{m}$, with a resolving power of $R \sim 1000$ for sources that fill the 0.2\arcsec-wide slit. Each target was observed using three-shutter slitlets with a nodding pattern. Each microshutter subtends $0.20\arcsec \times 0.46\arcsec$ on the sky. The effective exposure time per source per grating/filter combination was 7,000 seconds ($\sim 1.94$ hours). \added{The NIRSpec/MSA observations were conducted in August 2023.}

\begin{figure*}[!ht]
    \centering
    \includegraphics[width=1.0\textwidth]{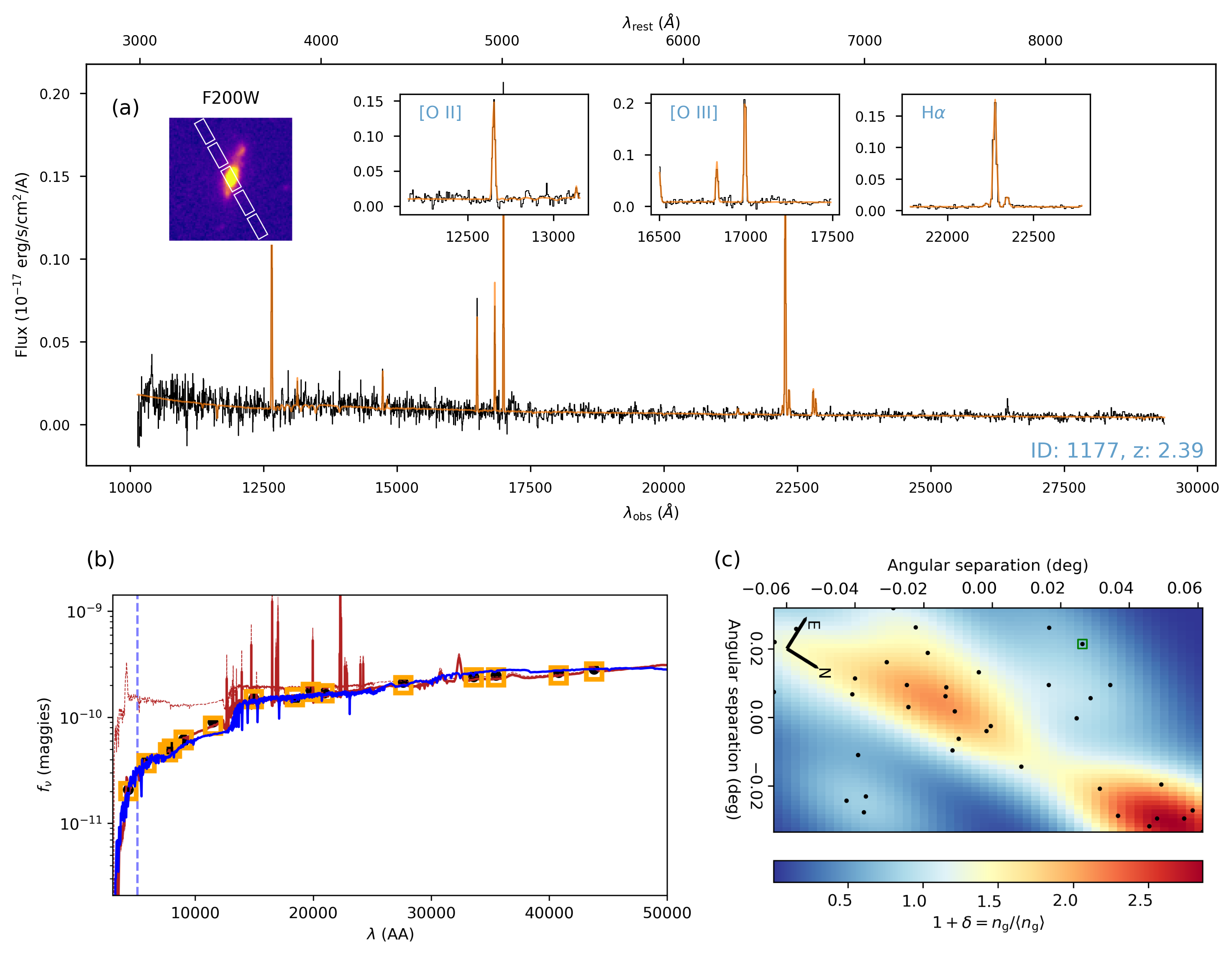}
    \caption{Example of spectroscopic and photometric data for galaxy SMILES-1177 at $z=2.4$. \textbf{(a)} medium-resolution MSA spectrum of the galaxy (black) with fitting from GELATO (orange). The insets from left to right show the F200W image with micro shutter placement, the zoom in on the [O II] emission line, the [O III] emission line, and the H$\alpha$ line, respectively. \textbf{(b)} photometric data from NIRCam and HST along with the best-fit SEDs from {\tt Prospector}. The red curve plots our best-fit SED model as decribed in Section \ref{sec:methods} while the blue curve plots an SED fit without nebular emission following the method in \citet{lyu_active_2024}. We find that the rest-UV SED fitting is not sensitive to the methods. The dashed red line plots the best-fit intrinsic SED, which is used to measure $L_{\rm UV}$ in this work. The vertical dashed line labels 1500\AA\ in the rest frame. \textbf{(c)} Photometric density map within the SMILES footprint. Black points mark JADES galaxies within 25 comoving-Mpc from the redshift of the target galaxy, which is highlighted in the green box.}
    \label{fig:spectra}
\end{figure*}

We reduced and calibrated the data using the {\tt JWST Calibration Pipeline} \citep{bushouse_jwst_2022}, version 1.14.0, with CRDS\footnote{Calibration Reference Data System: \url{https://jwst-crds.stsci.edu/}} context jwst\_1236.pmap. Briefly, we used the {\tt calwebb\_detector1} stage to create uncalibrated slope images from the ``uncal.fits'' files, followed by the {\tt calwebb\_spec2} stage to perform flat-fielding, flux calibration, and local background subtraction, producing resampled spectra for each nodding position. Finally, we performed spectral combination and extraction using {\tt calwebb\_spec3}. We also implemented custom scripts for additional hot pixel rejection, 1/f noise removal, and treatment for extended sources. Detailed data reduction procedures, reduced spectra, and the redshift catalog will be presented in a future data release (Y.~Zhu et al., in prep.). Figure \ref{fig:spectra} (a) displays an example of a processed 1D spectrum.

Since we need to measure $\xi_{\rm ion}$ using the H$\alpha$ luminosity from the spectra and $L_{\rm UV}$ from photometric data (see Section \ref{sec:methods}), it is necessary to apply a correction for slit loss. Slit loss occurs because the slits used in the MSA spectra only capture a portion of the extended galaxy's light, while photometry typically captures the full light profile of the galaxy. The default pipeline corrects for slit loss assuming a point source, but galaxies often have extended light profiles. Therefore, we need to account for the extended emission that is missed by the slits. To correct for this extended emission, we first measured the fraction of flux contained within the shutter aperture compared to the total flux from Kron-convolved photometry in several NIRCam bands. This step ensures that we capture the full light profile of the galaxy as observed in broad-band photometry. Next, we plotted this fraction against the central wavelength for each band and fitted a second-order polynomial to model the wavelength dependence of the slit loss.  \added{We find that the correction factor typically varies by about 20\% across the $1$–$3\ \mu$m range, though the wavelength dependence can vary significantly between sources.} We then applied this correction curve to the spectral flux. Finally, we tested the accuracy of our correction by comparing synthetic photometry generated from the corrected spectra with the actual photometric data. We found that this method yields a typical error of $\pm 0.2$ dex, with a median offset of 0, indicating that our correction procedure effectively compensates for the slit loss across the sample.

\added{Spectroscopic redshifts were determined from the observed spectra. Briefly, we visually identified prominent emission lines such as [O~{\sc ii}], H$\beta$, [O~{\sc iii}], H$\alpha$, and [N~{\sc ii}], following procedures similar to those described in \citet{deugenio_jades_2024}. Each spectrum was independently examined by at least two team members, and redshift measurements were validated using automated line-fitting tools such as GELATO (details in Y. Zhu et al. in prep.). Among the 168 observed objects, 151 have photometric redshifts $z_{\rm phot} > 0.3$ and are retained for analysis (photo-$z$ is adopted when spec-$z$ is unavailable or uncertain, as photo-$z$ is used for the overdensity measurements). }
\added{We exclude 24 AGNs (see below), 7 extended sources that suffered from severe self-subtraction, 3 sources outside or on the edge of the JADES GOODS-S NIRCam imaging footprint, and 9 spectra that are excessively noisy or strongly contaminated by neighboring objects, leaving 108 galaxies. We further remove 15 targets for which emission line fitting failed in GELATO, yielding a final sample of 93 galaxies. 
Of these, 67 have H$\alpha$ coverage (50 of which also have [O~{\sc iii}]), and an additional 12 galaxies have only [O~{\sc iii}] detected.} The spectra used in this work are listed in Table \ref{tab:sample}. \added{For $\xi_{\rm ion}$ measurements, we use the 67 galaxies at $1.0<z<3.1$ that have H$\alpha$ coverage.} We also included galaxies that have [O III] detections to compare their properties (see Section \ref{sec:discussion}). 

\added{We define an emission line as detected if the line flux exceeds three times the associated fitting error. All 67 galaxies used for $\xi_{\rm ion}$ analysis have H$\alpha$ detected at $>3\sigma$ significance. The 3$\sigma$ H$\alpha$ flux limit spans from $3.8 \times 10^{-18}$ to $1.1 \times 10^{-16}$ erg s$^{-1}$ cm$^{-2}$, with a median of $1.9 \times 10^{-17}$ erg s$^{-1}$ cm$^{-2}$. This corresponds to a minimum H$\alpha$ luminosity of $\log_{10}(L_{\rm H\alpha}/\text{erg s}^{-1}) = 52.76$. The derived SFRs from SED fitting span 0.0065–20.1 $M_\odot$/yr, and stellar masses span $\log_{10}(M_*/M_\odot) = 8.01$–$9.93$ (95\% ranges). We also detect 39 galaxies with both H$\alpha$ and H$\beta$ (for Balmer decrement), 32 galaxies with both [O~{\sc iii}] and [O~{\sc ii}] (for $O_{32}$), and 12 with only [O~{\sc ii}].}
\added{We also verify that the sample is broadly representative of the star-forming population. Based on SFR and stellar mass measurements from SED fitting, most galaxies lie within $\pm$1 dex of the star-forming main sequence at similar redshifts \citep[e.g.,][]{popesso_main_2023}, indicating that our sample spans a typical range of star formation activity. A small fraction of sources fall below the main sequence, possibly reflecting the presence of quiescent galaxy candidates with suppressed star formation rates.}

\added{In this work, we desired to exclude AGNs to the greatest extent possible. We therefore eliminated all those fulfilling the selection criteria in the comprehensive study of \citet{lyu_agn_2022, lyu_active_2024}, i.e., identified through: (1) SED fitting with JWST/SMILES and JADES photometry; (2) high ratios of X-ray to radio output; (3) high X-ray luminosities; (4) Chandra and/or HST variability; (5) pre-JWST optical or near infrared spectroscopy; (6) radio loudness; and (7) flat spectrum radio spectra. These selections are comprehensive; they include equally thoroughly the unobscured AGN as well as the obscured ones.}

\section{Methods}\label{sec:methods}
\subsection{SED fitting and UV luminosity}
To derive the physical properties of the galaxies, we use the {\tt Prospector} software \citep{johnson_stellar_2021} combined with the {\tt Parrot} artificial neural network (ANN) emulator\footnote{\url{https://github.com/elijahmathews/MathewsEtAl2023}} for acceleration \citep{mathews_as_2023} to fit the spectral energy distribution (SED) based on the multi-band Kron convolved photometry from JADES \citep{eisenstein_jades_2023,bunker_jades_2024,rieke_jades_2023,williams_jems_2023}. The photometric bands utilized include a combination of HST and JWST filters: the HST ACS bands (F435W, F606W, F775W, F814W, and F850LP); HST WFC3/IR bands (F105W, F125W, F140W, and F160W); and the JWST NIRCam bands (F090W, F115W, F150W, F182M, F200W, F210M, F277W, F335M, F356W, F410M, F430M, F444W, F460M, F480M). \added{We do not include MIRI bands in the SED fitting, as most galaxy properties for our $z \sim 1$–3 sample can be well constrained using HST and NIRCam photometry alone. We verified that including MIRI photometry does not significantly change the inferred parameters. Moreover, including MIRI would require convolution to a larger PSF, which we avoid for consistency. MIRI data were, however, used for AGN identification in \citet{lyu_active_2024} and any sources with notable AGN contribution have been removed from the sample.}

We adopt the assumptions and methodology outlined in \citet{mathews_as_2023}. Specifically, the physical model is based on the Prospector-$\alpha$ model  \citep{leja_how_2019}, which includes 14 parameters that quantify a galaxy's mass, stellar and gas-phase metallicity, star formation history (SFH), dust properties, etc. Modifications to the fiducial Prospector-$\alpha$ model extend the lower limit on the cumulative stellar mass formed down to $10^6$ $M_\odot$ for low-mass solutions at low redshift. Additionally, because the training data's photometric coverage includes the rest-frame mid-IR, the model allows three parameters governing the thermal dust emission (based on \citealp{draine_infrared_2007}) to vary: the PAH mass fraction ($Q_{\rm PAH}$), the minimum radiation field strength for dust emission ($U_{\rm min}$), and the fraction of starlight exposed to radiation fields. Furthermore, each galaxy is fitted with its spectroscopic redshift. This approach results in an 18-parameter physical model, where these parameters are used as inputs to the ANN emulators. For our fitting, we use the ``stitched model'' (see \citealp{mathews_as_2023} for details) with the redshift of each galaxy modeled as a truncated normal distribution centered at the spectroscopic redshift. We use a non-parametric star formation history (SFH) described by \citet{leja_how_2019}, modeled as seven SFR bins controlled by the continuity prior. Utilizing the ANN emulator greatly speeds up the fitting procedure, while introducing no apparent bias compared with \added{traditional} SPS codes with typical differences only of order 25\%–40\% for stellar mass, stellar metallicity, SFR, and stellar age \citep{mathews_as_2023}.The fitting accuracy is also confirmed by comparisons with our tests following the Prospector setup described in \citet{ji_jades_2023}, and \added{the inferred physical properties (e.g., stellar mass, SFR) only differ by $<0.15$ dex.} We also verified that the best-fit SEDs are broadly consistent with those from \citet{lyu_active_2024} in the rest-UV part. An example of the best-fit SED is shown in Figure \ref{fig:spectra} (b).

Once we acquire the SED for each galaxy, we measure the UV luminosity ($L_{\rm UV}$) by averaging the flux over $1450-1550$ \AA\ in the rest frame on the intrinsic (without dust attenuation) SED. We adopt the uncertainty from the nearest photometric data points. Additionally, we add a nominal  $\pm 30\%$ error to account for uncertainties and potential biases in the SED fitting procedure. Finally, we visually inspect the fitting results and remove $L_{\rm UV}$ values with failed SED fittings that significantly deviate from the rest-UV photometry. \added{There are 4 such cases in the final sample of 79 galaxies; for these, we replace the $L_{\rm UV}$ values from SED fitting with those manually measured from the photometry.}

\subsection{\texorpdfstring{$\xi_{\rm ion}$}{Lg} Measurements}
Once we have measured $L_{\rm UV}$, we can compute $\xi_{\rm ion}$ as 
\begin{equation}
    \xi_{\rm ion} = \dot{n}_{\rm ion} / L_{\rm UV}\,, 
\end{equation}
where $\dot{n}_{\rm ion}$ is the production rate of ionizing photons. H$\alpha$ serves as a good proxy for $\dot{n}_{\rm ion}$ \citep[see, e.g.,][]{osterbrock_astrophysics_2006}, with 
\begin{equation}
\dot{n}_{\rm ion} = 7.28 \times 10^{11} L(\text{H}\alpha)\,,    
\end{equation}
where $\dot{n}_{\rm ion}$ is in units of photons per second and $L(\text{H}\alpha)$ is in erg per second. Throughout this work, we assume no ionizing photon escape fraction, i.e., $f_{\rm esc} = 0$. A non-zero $f_{\rm esc}$ would introduce a factor of $(1 - f_{\rm esc})$ in the inferred $\dot{n}_{\rm ion}$, resulting in an underestimation of the intrinsic $\xi_{\rm ion}$. Additionally, we assume case B recombination with a temperature of $10^4$ K \citep[following e.g., ][]{simmonds_low-mass_2024,pahl_spectroscopic_2024}.

On the slit-loss-corrected spectra (see Section \ref{sec:data}), we further corrected the flux for dust attenuation. To ensure consistency with the $L_{\rm UV}$ measurements, we derived the attenuation curve by computing the ratio between the intrinsic SED and the best-fit observed SED from our SED fitting procedure. Additionally, we verified that using an alternative method, such as the Balmer decrement, does not significantly affect our conclusions (see Appendix \ref{sec:balmer}).

We then fit the corrected spectra using the GELATO software\footnote{\url{https://github.com/TheSkyentist/GELATO}} \citep[][]{hviding_theskyentistgelato_2022}. GELATO provides comprehensive support for both stellar population synthesis (SSP) continuum modeling and emission line fitting. The emission lines in GELATO are modeled as Gaussians, with parameters for redshift, flux, and dispersion. The continuum is constructed using the Extended MILES stellar library (E-MILES) SSP models,\footnote{\url{http://research.iac.es/proyecto/miles/}} assuming a Chabrier initial mass function (Chabrier IMF; \citealp{chabrier_galactic_2003}) and spanning a range of representative metallicities and ages \citep{hviding_new_2022}. 

\added{We fit major emission lines, including H$\alpha$, H$\beta$, H$\gamma$, [O~{\sc i}], [O~{\sc ii}], [O~{\sc iii}], [S~{\sc ii}], and [N~{\sc ii}]. The redshift prior is given the spectroscopic redshift measured in Section~\ref{sec:data}. For line fluxes, we impose constraints only on doublets with known intrinsic ratios (e.g., [O~{\sc iii}] $\lambda4960/\lambda5008 = 1/3$). We allow an additional broad component for H$\alpha$ or [O~{\sc iii}] when the line width exceeds 500 km/s, and also allow for shifted ``outflow'' components for [O~{\sc iii}]. Although GELATO performs simultaneous continuum and line fitting, it does not return continuum parameters in its output files, preventing a direct comparison to SED fitting from Prospector. We assess the quality of fits based on the reported flux errors and visual inspection, and remove 15 galaxies from our analysis due to poor fits likely caused by contamination or low S/N. We also verified that GELATO produces consistent line fluxes compared to those from pPXF \citep{cappellari_ppxf_2012} fits. A complete emission line catalog will be presented in Zhu et al. (in prep.).}

\begin{figure*}[!ht]
    \centering
    \includegraphics[width=0.95\textwidth]{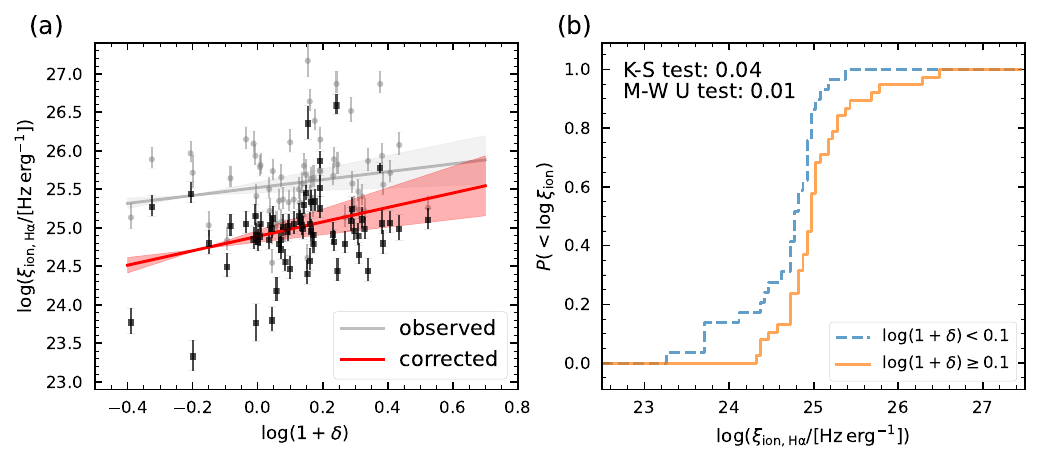}
    \caption{
    \textbf{(a)} \added{\textbf{Trend}} between ionizing photon production efficiency and galaxy overdensity. Black and gray symbols plot results after and before the dust attenuation correction, respectively. The red line plots the best linear fit between $\log\xi_{\rm ion, H\alpha}$ and $\log( 1+\delta)$, with the shaded region showing $68\%$ range of the slope. The linear relation is statistically significant with $p$-value of 0.01 (see text for details). The gray line and shaded region plot the fitting for results without dust attenuation correction for reference \added{( $\log (\xi_{\rm ion} / \rm [Hz\,erg^{-1}]) = 0.51_{-0.35}^{+0.35} \log(1+\delta) + 25.52_{-0.07}^{+0.07}$)}. (b) Cumulative distribution function of $\log\xi_{\rm ion, H\alpha}$ for galaxies in high density $\log(1+\delta) > 0.1$ and low density $\log( 1+\delta) < 0.1$ regions, respectively. The results suggest that galaxies in high density regions produce ionizing photons more efficiently (with a $p$-value of 0.01 based on Mann–Whitney U test).}
    \label{fig:xion_overdensity}
\end{figure*}

\subsection{Galaxy Overdensity Measurements}
To compute galaxy overdensities within the SMILES footprint, we utilize the JADES photometric catalog \citep{eisenstein_overview_2023,rieke_jades_2023}. We begin by filtering the catalog to select galaxies with $0.1 < z < 7$ and apparent magnitudes corresponding to $1$ $\mu$m in the rest frame $m_{\rm 1 \mu m}< 29$. We use photometric redshift values from \citet{hainline_cosmos_2024} based on EAZY photo-z algorithm \citep{brammer_eazy_2008}, as well as their 16th and 84th percentiles, $z_{16}$ and $z_{84}$, respectively. The SMILES target galaxies are identified by matching their IDs to the JADES catalog. We then calculate the spatial distribution of galaxies within a rotated coordinate system, centered on the SMILES MSA field, and apply a bounding box to define the region of interest. To avoid edge effects in our density estimation, we mirror galaxies within the bounding box to artificially extend the dataset.

For each target galaxy, we compute the galaxy overdensity by slicing the data into comoving distance bins of 25 Mpc. Within each bin, we employ a two-dimensional Gaussian Kernel Density Estimation (KDE) to estimate the projected galaxy density in RA–Dec. The bandwidth of the KDE is optimized using cross-validation. The overdensity $1+\delta = n_{\rm g} / \langle n_{\rm g} \rangle$ is then defined as the ratio of the KDE value at a galaxy's \added{location (i.e., the local projected density) to the mean KDE value across the full spatial extent of the SMILES field in that redshift slice.}  Additionally, we weight each galaxy's contribution to the density estimation by its redshift uncertainty, calculated from the Gaussian distribution of $z_{16}$ and $z_{84}$. 
The final overdensity value for each SMILES target galaxy is derived by averaging the densities across all comoving distance bins in which it appears, using the maximum-weighted density estimate. \added{While the KDE provides a \textit{relative} measure of local density, we do not assign formal uncertainties to individual overdensity values, consistent with similar studies (e.g., \citealt{chartab_large-scale_2020,helton_jwst_2023}).} In Appendix \ref{app:overdensity}, we also tested that using the overdensity measurements based on CANDELS data \citep{chartab_large-scale_2020} does not change our conclusions. 

Figure \ref{fig:overdensity-z} displays the distribution of overdensity and redshift of galaxies studied in this work. The majority of galaxies have $z<3.7$ and are used for direct $\xi_{\rm ion}$ measurements. Overall, the overdensity distribution is roughly symmetric about $\log (1+\delta)  = 0.1$. \added{We adopt this threshold to divide the sample into low- and high-density regions in Section \ref{sec:discussion}, although it is not physically motivated.} \added{This threshold reflects the density distribution in the SMILES field and the selection toward relatively massive galaxies, which tend to live in massive halos and slightly overdense regions.} There is no strong correlation between $\log (1+\delta)$ and $z$ in our sample, which assures that the \added{trend} between $\xi_{\rm ion}$ and $\log (1+\delta)$ (Section \ref{sec:discussion}) is not caused, or at least dominated by, redshift evolution found in e.g., \citet{pahl_spectroscopic_2024}. \added{We note that our results are based on a relatively small field ($\sim$34 arcmin$^2$) and limited dynamic range in overdensity. Therefore, future observations over wider areas and a broader range of environments are essential to confirm and strengthen the observed trends.}

\begin{figure*}[!ht]
    \centering
    \includegraphics[width=1\textwidth]{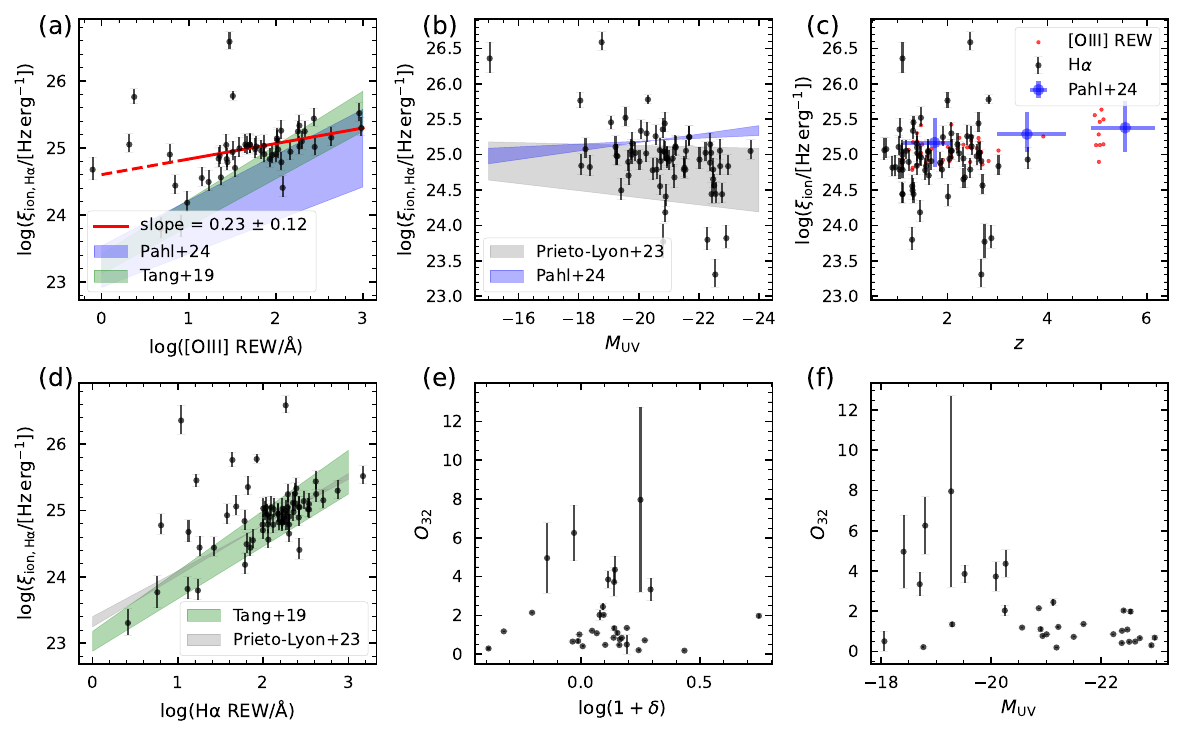}
    \caption{Other galaxy properties explored in this work. 
    \textbf{(a)} The correlation between $\log\xi_{\rm ion, H\alpha}$ and $\log$([OIII] REW). The red line shows the best-fit linear regression with a slope of $0.23 \pm 0.12$ for $\log$([OIII] REW)$>1$, and the dashed line shows the extrapolation. The shaded regions represent the results from \citet{pahl_spectroscopic_2024} and \citet{tang_mmtmmirs_2019}.
    \textbf{(b)} The correlation between $\log\xi_{\rm ion, H\alpha}$ and $M_{\rm UV}$. Shaded regions indicate the results from \citet{prieto-lyon_production_2023} and \citet{pahl_spectroscopic_2024}.
    \textbf{(c)} The evolution of $\log\xi_{\rm ion}$ with redshift. Black data points show $\log\xi_{\rm ion}$ derived from the H$\alpha$ luminosity, and red data points mark those derived from [O III] REW.
    Blue error bars show the composite results from \citet{pahl_spectroscopic_2024}.
    \textbf{(d)} The correlation between $\log\xi_{\rm ion, H\alpha}$ and $\log$(H$\alpha$ REW). Shaded regions correspond to results from \citet{tang_mmtmmirs_2019} and \citet{prieto-lyon_production_2023}.
    \textbf{(e)} The relation between O$_{32}$ and galaxy overdensity, $\log(1+\delta)$, and no significant correlation is identified.
    \textbf{(f)} The slight anti-correlation between O$_{32}$ and $M_{\rm UV}$.
    }
    \label{fig:other_corr}
\end{figure*}

\section{Results and Discussion} \label{sec:discussion}

\subsection{Trend between \texorpdfstring{$\xi_{\rm ion}$}{Lg} and galaxy overdensity}

Figure \ref{fig:xion_overdensity} (a) displays our main result, the \added{trend} between $\xi_{\rm ion}$ and galaxy overdensity for 67 galaxies at \added{$1.0 < z < 3.1$}. We perform least-squares linear fitting using the {\tt scipy} package based on 1,000 bootstrap resamplings of the data points. We find a strong positive \added{trend}:
\begin{equation}
    \log (\xi_{\rm ion} / \rm [Hz\,erg^{-1}]) = 0.94_{-0.46}^{+0.46} \log(1+\delta) + 24.89_{-0.08}^{+0.08} \,.
\end{equation}
We performed a Wald test \added{implemented in {\tt scipy.stats.linregress}} \citep[][]{virtanen_scipy_2020} with t-distribution and find $p = 0.01$. 
This $p$-value of $<0.05$ indicates that the positive \added{trend} is statistically significant ($\sim 2\sigma$ level). 
\added{We note that substantial scatter exists, and a few galaxies lie significantly above or below this trend. We have checked their spectral fitting and found no issues. These outliers may reflect galaxies with varying dust attenuation, bursty star formation, or recent stochastic variations in star formation history. Still, we caution that this result should be interpreted as an environmental trend rather than a strict linear correlation.}

If we divide the sample into high-density and low-density subsamples, with $\log(1+\delta) > 0.1$ and $\log(1+\delta) < 0.1$ (the median value in logarithmic space, corresponding to 1.25 times the mean density in linear space), we observe a difference in the distribution of $\xi_{\rm ion}$. Figure \ref{fig:xion_overdensity} (b) shows the cumulative distribution function (CDF) of $\xi_{\rm ion}$ for the two cases. We see that the high-density curve is shifted to the right compared to the low-density subsample. A Mann-Whitney U test returns a $p$-value of 0.01, which suggests that the difference in distributions is statistically significant. We chose the Mann-Whitney U test instead of the Kolmogorov-Smirnov (K-S) test because the K-S test is more sensitive to differences in the middle of the distribution rather than the tails. Still, the K-S test yields a $p$-value of 0.04, and is consistent with the Mann-Whitney U test. 
Furthermore, we find a mean ratio of 2.43 (median ratio of 1.58) between $\xi_{\rm ion}$ in high-density versus low-density environments.

\added{
These results imply that galaxies in high-density regions tend to produce ionizing photons more efficiently. A burstier star-formation history in overdense environments may contribute to this trend. By construction, our measurement of $\xi_{\rm ion}$ compares the H$\alpha$ luminosity, which traces star formation over the past $\sim$10 Myr, to the UV continuum luminosity, which reflects star formation averaged over longer timescales of $\sim$100 Myr. Therefore, galaxies with recent starbursts relative to their long-term average will exhibit elevated $\xi_{\rm ion}$ values. Such bursts may be more frequent in dense environments, where environmental effects such as galaxy-galaxy interactions, mergers, or inflows of cold gas can trigger enhanced star formation activity \citep[e.g.,][]{elbaz_reversal_2007,koyama_massive_2013}. 

Consistent with this picture, observations at cosmic noon have revealed that galaxies in overdensities tend to show enhanced H$\alpha$ luminosities compared to the field, as seen in the environmental dependence of the H$\alpha$ luminosity function \citep{sobral_dependence_2011}. Additionally, red and dusty H$\alpha$ emitters have been identified in group-like environments \citep[e.g.,][]{garn_obscured_2010}, potentially indicating rejuvenated star formation in massive galaxies residing in dense regions. These combined effects could result in a systematically higher ionizing photon output per unit UV luminosity.

However, we caution that our current field of view is relatively small, which limits our ability to account for cosmic variance. In addition, the slit-loss correction for multi-object spectroscopy remains non-trivial, and there is no consensus on the optimal method for MSA spectra. Dust attenuation corrections for high-redshift galaxies also introduce uncertainties. Nevertheless, we emphasize that even when omitting dust correction and using $L_{\rm UV}$ directly from the photometry instead of the best-fit SEDs, the observed trend between $\xi_{\rm ion}$ and overdensity remains. We also tested that removing the slit-loss correction, or applying a simplified linear correction, does not qualitatively change the result.
}

\subsection{Correlations with other emission line properties}
In this subsection, we explore several commonly studied emission line properties related to the ionization properties of galaxies as discussed in the literature. The results are displayed in Figure \ref{fig:other_corr}. 
\added{
To better contextualize our comparisons with previous studies, we summarize the sample selection and methodology of each work. 
\citet{simmonds_ionising_2024} analyzed $z \sim 4$--9 galaxies in JADES using deep photometry and SED fitting. $\xi_{\rm ion}$ was derived from modeled H$\alpha$ luminosities and rest-UV continuum, spanning $-21 < M_{\rm UV} < -16$. 
\citet{pahl_spectroscopic_2024} studied $1.06 < z < 6.71$ galaxies in CEERS and JADES with NIRSpec, deriving $\xi_{\rm ion}$ both from direct H$\alpha$ measurements and from empirical correlations with [O~III] REW, over $-22 < M_{\rm UV} < -17$.
\citet{tang_mmtmmirs_2019} used MMT/MMIRS to analyze bright galaxies at $1.3 < z < 2.4$, selected from ground-based observations, focusing on extreme emission line sources. 
\citet{prieto-lyon_production_2023} examined 370 galaxies at $z \sim 3$--7 from the GLASS and UNCOVER JWST surveys, spanning $-23 < M_{\rm UV} < -15.5$ (median $M_{\rm UV} \approx -18$), and derived $\xi_{\rm ion}$ using deep HST and JWST/NIRCam photometry through SED modeling.

While differences in sample selection and methodology exist, our $\xi_{\rm ion}$ derivation—based on spectroscopic H$\alpha$ luminosities and rest-UV photometry—is most comparable to the direct approaches used in \citet{pahl_spectroscopic_2024}.
}

Figure \ref{fig:other_corr} (a) shows the correlation between $\xi_{\rm ion}$ and [O III] rest-frame equivalent width (REW). As previously found in other studies \citep[e.g.,][]{tang_mmtmmirs_2019, boyett_extreme_2024,pahl_spectroscopic_2024}, [O III] REW can be tightly correlated with $\xi_{\rm ion}$. We perform a linear fit to the sample with $\log$([O III] REW/\rm \AA) $> 1$ and find a slope of $0.23 \pm 0.12$. However, we note that our data points lie above the literature relations. Possible reasons for this discrepancy include differences in sample selection (see e.g., \citealp{laseter_efficient_2024}) or variations in slit-loss and dust correction treatments.

Figure \ref{fig:other_corr}~(b) plots $\xi_{\rm ion}$ as a function of $M_{\rm UV}$. \added{Our results suggest a mild trend in which UV-faint galaxies (i.e., those with larger $M_{\rm UV}$) tend to exhibit higher $\xi_{\rm ion}$.} \added{This is broadly consistent with findings from several recent studies. For instance, \citet{castellano_ionizing_2023}, \citet{llerena_ionizing_2024}, and \citet{rinaldi_midis_2024} all report higher $\xi_{\rm ion}$ in UV-faint galaxies, particularly at $z > 4$. Similarly, \citet{simmonds_ionising_2024} find a comparable trend in JADES galaxies.} \added{In contrast, \citet{pahl_spectroscopic_2024} report a weak trend in the opposite direction, where UV-brighter galaxies show slightly higher $\xi_{\rm ion}$, while \citet{prieto-lyon_production_2023} also find UV-faint galaxies to be more ionizing efficient.} \added{Meanwhile, \citet{shivaei_mosdef_2018} and \citet{bouwens_lyman-continuum_2016} find no strong correlation, although their samples are mostly limited to UV-bright systems ($M_{\rm UV} \gtrsim -20$).} \added{These differences across studies underscore the need for larger and more uniformly selected samples to better establish the underlying trends between $\xi_{\rm ion}$ and UV luminosity.}

In Figure \ref{fig:other_corr} (c), we plot $\xi_{\rm ion}$ as a function of redshift, along with values derived from [O III] REW, based on the correlation acquired in Figure \ref{fig:other_corr} (a). Our results are consistent with those of \citet{pahl_spectroscopic_2024} and confirm the mild redshift evolution found in their work.
In Figure \ref{fig:other_corr} (c), we plot $\xi_{\rm ion}$ as a function of redshift, along with values derived from [O~III] REW, based on the correlation acquired in Figure \ref{fig:other_corr} (a). Our results are consistent with those of \citet{pahl_spectroscopic_2024} and confirm the mild redshift evolution found in their work. \added{This mild upward trend in $\xi_{\rm ion}$ with redshift is also consistent with findings from \citet{shivaei_mosdef_2018}, \citet{castellano_ionizing_2023}, and \citet{llerena_ionizing_2024}. Together, these studies support the interpretation that galaxies at higher redshifts tend to be more efficient producers of ionizing photons, possibly due to lower metallicities and more intense star formation activity \citep[e.g.,][]{laseter_efficient_2024}.}

Figure \ref{fig:other_corr}~(d) shows the correlation between $\xi_{\rm ion}$ and H$\alpha$ REW. We find a strong positive correlation, with our data points being broadly consistent with the results of \citet{tang_mmtmmirs_2019}. \added{Such a positive correlation has also been identified across a wide redshift range in other studies \citep[e.g.,][]{prieto-lyon_production_2023,rinaldi_midis_2024}, suggesting that the relation does not strongly depend on redshift evolution.} \added{This trend is also physically expected: both $\xi_{\rm ion}$ and H$\alpha$ REW are sensitive to the age of the stellar population, with younger stellar populations producing higher values of both quantities, as shown in BPASS-based models (e.g., Prieto-Jiménez et al.\ 2025, submitted).} \added{Additionally, $\xi_{\rm ion}$ depends on the presence of massive stars (O and B types), which are affected by the IMF and stochasticity in their formation; extreme values of $\xi_{\rm ion}$ ($>25.5$) may only be reached in bursts with enhanced massive star formation \citep[e.g.,][]{stanway_stellar_2016}.}

Finally, we examine O$_{32}$, a potential tracer of the escape fraction of ionizing photons \citep[e.g.,][]{chisholm_accurately_2018,choustikov_physics_2024}, in Figure \ref{fig:other_corr} (e) and (f). We do not observe a strong correlation between O$_{32}$ and galaxy overdensity in our data, as shown in Figure \ref{fig:other_corr} (e). However, Figure \ref{fig:other_corr} (f) reveals a mild anti-correlation between O$_{32}$ and $M_{\rm UV}$. These findings suggest that while O$_{32}$ may not be strongly influenced by the local environment, it does exhibit a dependence on UV luminosity, consistent with other studies \citep[e.g.,][]{pahl_spectroscopic_2024}. 
\added{We note that O$_{32}$ may also be affected by other interstellar medium (ISM) conditions beyond $f_{\rm esc}$. Recent works suggest that more accurate (albeit indirect) predictions of the escape fraction can be achieved by combining multiple galaxy properties beyond just O$_{32}$ \citep[e.g.,][]{choustikov_physics_2024}.} 
\added{Previous studies have investigated the correlation between $\xi_{\rm ion}$ and O$_{32}$: \citet{shen_ngdeep_2025} and \citet{llerena_ionizing_2024} found a positive correlation, while \citet{shivaei_mosdef_2018} reported a weaker or less significant relation.} 
\added{Nevertheless, $f_{\rm esc}$ may still be linked to the environment, as suggested by observations of ionized bubbles in high-density regions \citep[e.g.,][]{saxena_jades_2024,witstok_inside_2024}.}

Since there is no strong correlation between O$_{32}$ and galaxy overdensity, and our sample does not exhibit a strong dependence of $M_{\rm UV}$ on overdensity \added{(see Figure \ref{fig:MUV_density})}, the only significant correlation that remains is between $\xi_{\rm ion}$ and overdensity. Therefore, in our sample, the total ionizing photon production rate, $\dot{N}$, is likely to be positively correlated with overdensity, at least in a first-order estimation. \footnote{By performing a linear regression between $\dot{n}_{\rm ion}$ and $\log(1+\delta)$, we find a slop of $0.51\pm 0.32$ with a $p$-value of 0.12. If we assume a constant escape fraction, then $\dot{N}$ would be positively correlated with $\log(1+\delta)$, although not statistically significant.}
This implies that galaxies in overdense regions may contribute more efficiently to the reionization process. However, overdense regions may also have higher gas densities and shorter mean free paths for ionizing photons, which complicates the overall effect on the IGM. As a result, while galaxies in overdensities may produce ionizing photons more efficiently, the net impact on reionization remains uncertain and requires further investigation on the escape fraction and the mean free path of ionizing photons.

\section{Summary}\label{sec:summary}

In this work, we investigated the ionizing photon production efficiency ($\xi_{\rm ion}$) and its relation to galaxy overdensity, along with other emission line properties, using a sample of \added{79} galaxies from the SMILES survey. We found a positive \added{trend} between $\xi_{\rm ion}$ and galaxy overdensity for 67 galaxies with H$\alpha$ detections at \added{$1.0 < z < 3.1$}. The slope of this \added{trend}, $0.94_{-0.46}^{+0.46}$, has a p-value of 0.01, indicating that it is statistically significant. The result suggests that galaxies in high-density regions produce ionizing photons more efficiently than those in lower-density environments, \added{possibly consistent with the observed burstier star-formation histories in overdense environments}.

When dividing the sample into high-density and low-density subsamples, we observed a statistically significant difference in the distributions of $\xi_{\rm ion}$, with a mean ratio of 2.43 between the two regions. Additionally, we explored correlations between $\xi_{\rm ion}$ and other emission line properties. We confirmed a positive correlation with [O III] REW, although our data points were systematically higher than those found in the literature \citep[e.g.,][]{tang_mmtmmirs_2019,pahl_spectroscopic_2024}. A mild redshift evolution of $\xi_{\rm ion}$ was also observed, consistent with \citet{pahl_spectroscopic_2024}. Furthermore, we examined O$_{32}$, a potential tracer of the escape fraction of ionizing photons, and found no significant correlation between O$_{32}$ and galaxy overdensity, but a mild anti-correlation with $M_{\rm UV}$ was evident. Given the absence of strong correlations between O$_{32}$ and overdensity, and between $M_{\rm UV}$ and overdensity, the positive \added{trend} between $\xi_{\rm ion}$ and overdensity implies that the total ionizing photon production rate ($\dot{N}$) is likely higher in overdense regions. \added{We also note that our analysis assumes zero escape fraction ($f_{\rm esc} = 0$), meaning all ionizing photons contribute to nebular emission. If $f_{\rm esc}$ is non-zero, the inferred $\xi_{\rm ion}$ would be underestimated. While  $f_{\rm esc}$ may depend on galaxy properties, any dependence on large-scale environment like overdensity remains uncertain.}

Our results represent the first measurement of the \added{trend} between $\xi_{\rm ion}$ and galaxy overdensity, highlighting the potential role of overdense regions in driving reionization. 
In addition, they are consistent with the picture that the star forming properties in overdensities may differ systematically from those for field galaxies (see e.g., \citealp{hayashi_properties_2011,harikane_silverrush_2019}). If our $\log\xi_{\rm ion}$-$\log(1+\delta)$ \added{trend} still holds at $z\sim 6$, then our discovery can naturally explain the higher transmitted flux observed in the \lya\ forest near overdensities traced by [O III] emitters during the epoch of reionization \citep[e.g.,][]{jin_spectroscopic_2024}. Nevertheless, future observations with a larger survey field are needed to address the issue of cosmic variance.  Additionally, future modeling work may help address uncertainties from slit-loss corrections and dust attenuation for MSA observations of high-$z$ galaxies.

\begin{acknowledgements}
\added{We thank the anonymous reviewer for their helpful comments.}
YZ, YS, ZJ, NB, MJR, and CNAW acknowledge support from the NIRCam Science Team contract to the University of Arizona, NAS5-02015. 
SA, JL, GHR and JM acknowledge support from the JWST Mid-Infrared Instrument (MIRI) Science Team Lead, grant 80NSSC18K0555, from NASA Goddard Space Flight Center to the University of Arizona.
AJB acknowledges funding from the ``FirstGalaxies'' Advanced Grant from the European Research Council (ERC) under the European Union’s Horizon 2020 research and innovation programme (Grant agreement No. 789056).

This work is based on observations made with the NASA/ESA/CSA James Webb Space Telescope. The data were obtained from the Mikulski Archive for Space Telescopes at the Space Telescope Science Institute, which is operated by the Association of Universities for Research in Astronomy, Inc., under NASA contract NAS 5-03127 for JWST. These observations are associated with PID 1207, 1180, and 1963. The specific observations analyzed can be accessed via 
\dataset[https://doi.org/10.17909/et3f-zd57]{https://doi.org/10.17909/et3f-zd57}, \dataset[https://doi.org/10.17909/8tdj-8n28]{https://doi.org/10.17909/8tdj-8n28},  and \dataset[https://dx.doi.org/10.17909/fsc4-dt61]{https://dx.doi.org/10.17909/fsc4-dt61}
\added{\citep{rieke_george_systematic_2024,rieke_marcia_data_2024,williams_christina_data_2023}}. 

We respectfully acknowledge the University of Arizona is on the land and territories of Indigenous peoples. Today, Arizona is home to 22 federally recognized tribes, with Tucson being home to the O’odham and the Yaqui. The University strives to build sustainable relationships with sovereign Native Nations and Indigenous communities through education offerings, partnerships, and community service.

\added{This manuscript benefited from grammar checking and proofreading using ChatGPT (OpenAI; \url{https://openai.com/chatgpt}).}
\end{acknowledgements}

\begin{contribution}
\added{
    YZ led the data calibration and analysis, and wrote the initial draft of the paper. SA, JL, JM, GHR, NB, and IS contributed to the design and implementation of the observations. SA, JL, JM, and GHR also contributed to data calibration and AGN selection. YS contributed to visual inspection of the spectra and emission line fitting. All authors contributed to the interpretation of the results and the scientific discussion.
    }
\end{contribution}

\vspace{5mm}
\facilities{JWST, MAST}

\software{
{\tt astropy} \citep{astropy_collaboration_astropy_2022},
{\tt GELATO} \citep{hviding_new_2022,hviding_theskyentistgelato_2022}, 
{\tt JWST Calibration Pipeline} \citep{bushouse_jwst_2022},
{\tt scipy} \citep{virtanen_scipy_2020}
}

\appendix

\renewcommand{\thefigure}{A\arabic{figure}}
\setcounter{figure}{0}

\section{Results Based on Overdensity Measurements from Literature} \label{app:overdensity}

\begin{figure}
    \centering
    \includegraphics[width=1.0\linewidth]{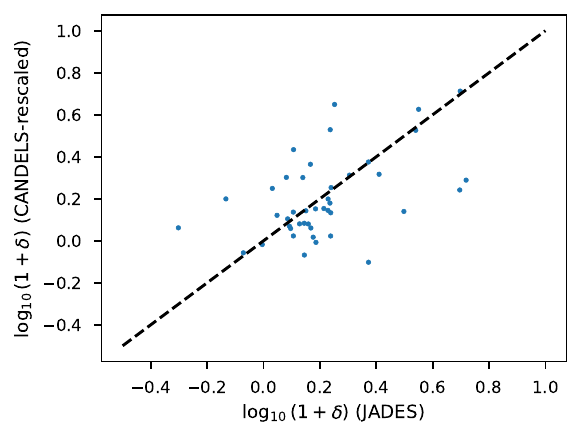}
    \caption{Comparison between overdensities measured in \citet{chartab_large-scale_2020} and this work. A magnitude cut of $m_{\rm F150W} < 26$ is applied to JADES data, and the density contrast values have been rescaled to match the scale of our measurements.}
    \label{fig:overdensity-cmp}
\end{figure}

\begin{figure}
    \centering
    \includegraphics[width=1.0\linewidth]{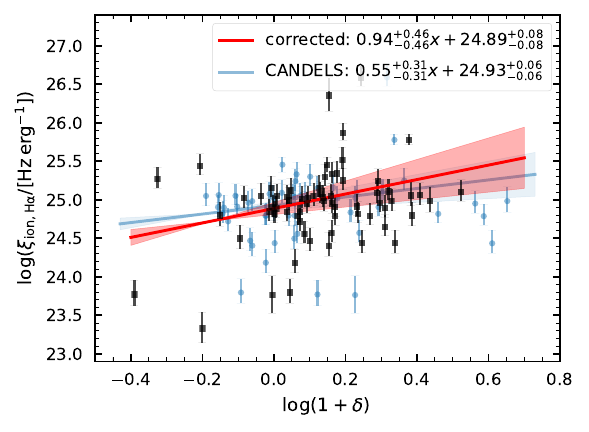}
    \caption{Similar to Figure \ref{fig:xion_overdensity} (a), but using rescaled overdensity measurements based on CANDELS (blue data points) from \citet{chartab_large-scale_2020}. Our main results (black data points and the red shaded region) are included as a reference baseline.}
    \label{fig:enter-label}
\end{figure}

To test the robustness of our overdensity measurement methods, we apply a different magnitude cut of $m_{\rm F150W} < 26$ to mimic the threshold used in \citet{chartab_large-scale_2020} for CANDELS data ($m_{\rm F160W} < 26$). The results are shown in Figure \ref{fig:overdensity-cmp}. The density contrast values from \citet{chartab_large-scale_2020} have been rescaled to match our values for display purposes. We find broad consistency between the two measurements, although some scatter is present. Additionally, when using the CANDELS measurements, we find that the positive \added{trend} between $\xi_{\rm ion}$ and $\log(1+\delta)$ is still present, with a slope of $0.55_{-0.31}^{+0.31}$, which is consistent with our main result within the $1\sigma$ level. Therefore, our density measurements and the $\xi_{\rm ion} - \log(1+\delta)$ \added{trend} measurements are robust.

\section{Results Based on Balmer Decrement} \label{sec:balmer}

\renewcommand{\thefigure}{B\arabic{figure}}
\setcounter{figure}{0}

\begin{figure}
    \centering
    \includegraphics[width=1.0\linewidth]{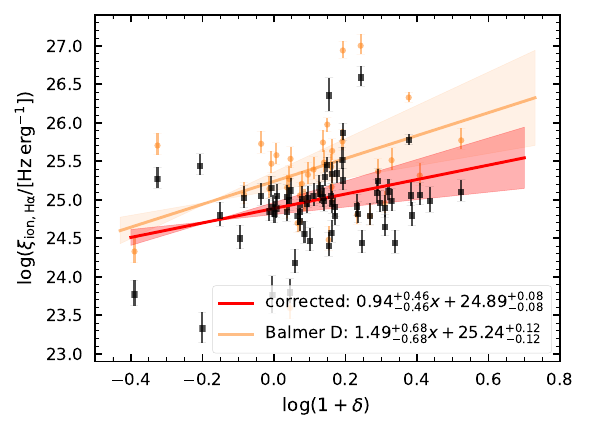}
    \caption{\added{trend} between $\log\xi_{\rm ion, H\alpha}$ and $\log(1 + \delta)$ using H$\alpha$ luminosities corrected by the Balmer decrement. The red line corresponds to the fit using our primary dust correction method, while the orange line represents the results using the Balmer decrement correction. Shaded regions indicate the 68\% confidence intervals.}
    \label{fig:balmer}
\end{figure}

The Balmer decrement corrected H$\alpha$ luminosity was obtained using the Milky Way (MW) extinction law, specifically the Cardelli-Clayton-Mathis (CCM) law with $R_V = 3.1$ \citep[][]{cardelli_relationship_1989}. Figure \ref{fig:balmer} shows the results, which also yield a positive \added{trend} $\log \xi_{\rm ion} = 1.49_{-0.68}^{+0.68} \log(1 + \delta) + 25.24_{-0.12}^{+0.12}$. We note that only 39 spectra are used to fit the relation here because not all spectra cover both the H$\alpha$ and H$\beta$ lines with detections.

\section{Potential Selection Effects}
\renewcommand{\thefigure}{C\arabic{figure}}
\setcounter{figure}{0}

\begin{figure*}
    \centering
    \gridline{
        \fig{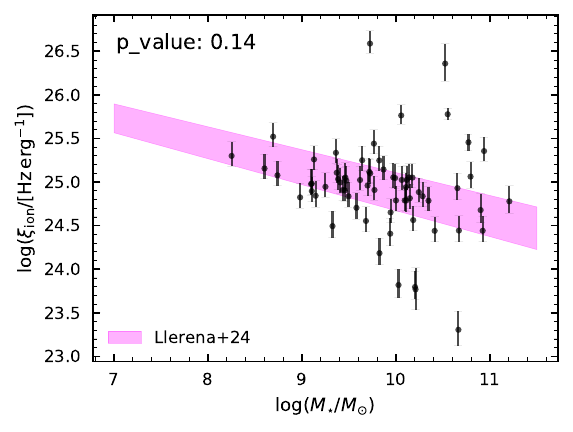}{0.45\linewidth}{(a)}
        \fig{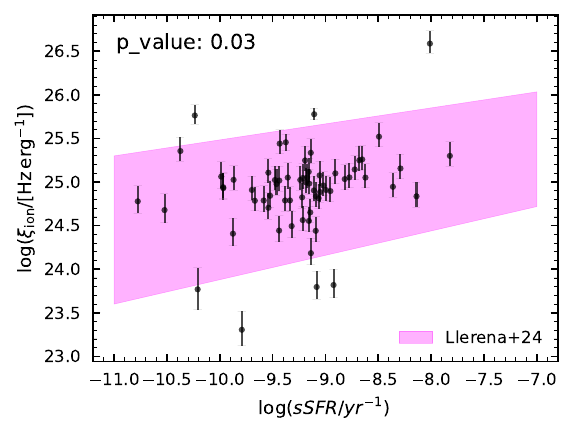}{0.45\linewidth}{(b)}
    }
    \gridline{
        \fig{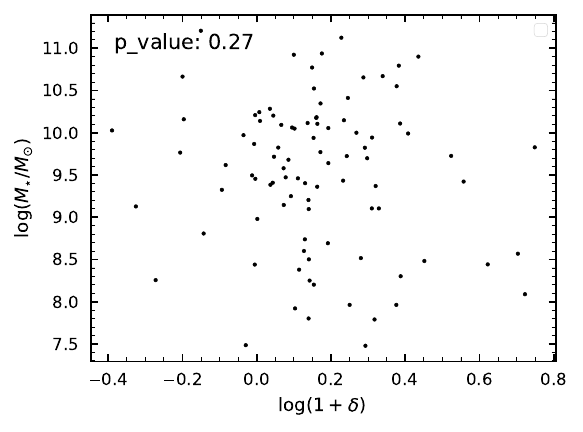}{0.45\linewidth}{(c)}
        \fig{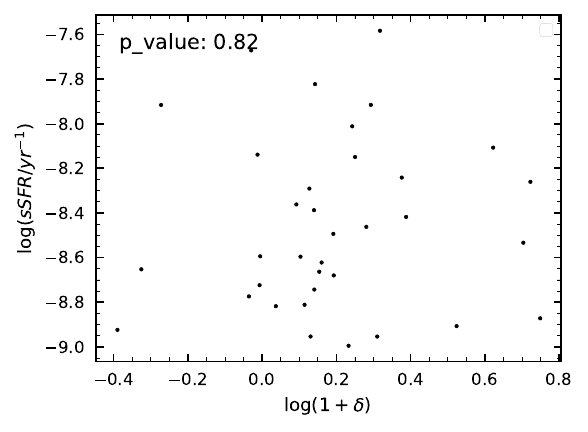}{0.45\linewidth}{(d)}
    }
    \caption{\added{Tests for potential selection effects. 
    \textbf{(a)} $\xi_{\rm ion}$ as a function of stellar mass. We observe a mild negative trend ($p = 0.14$), consistent with \citet{llerena_ionizing_2024}. 
    \textbf{(b)} $\xi_{\rm ion}$ versus specific star formation rate (sSFR), showing a positive trend ($p = 0.03$), also consistent with literature. 
    \textbf{(c)} Stellar mass versus galaxy overdensity, and 
    \textbf{(d)} sSFR versus galaxy overdensity. No significant correlations are found in (c) and (d), suggesting that the observed $\xi_{\rm ion}$–overdensity trend is unlikely to be driven by variations in mass or sSFR. Note that the shaded regions in (a) and (b) from \citet{llerena_ionizing_2024} show fitted trends but do not fully cover the scatter in their data.}}
    \label{fig:xion_M_sSFR}
\end{figure*}

\begin{figure}
    \centering
    \includegraphics[width=1.0\linewidth]{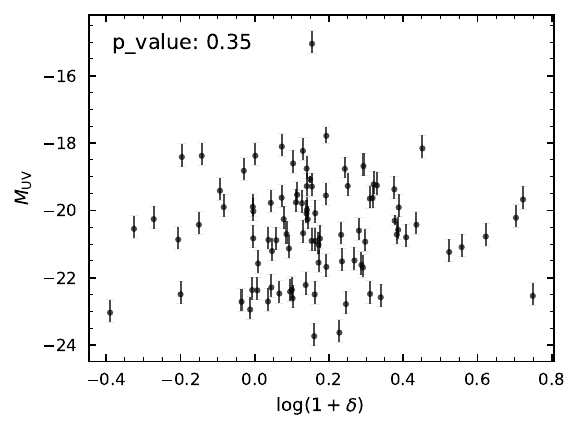}
    \caption{\added{Relation between $M_{\rm UV}$ and galaxy overdensity $\log(1 + \delta)$. No significant correlation is found, with a $p$-value of 0.35.}}
    \label{fig:MUV_density}
\end{figure}

\added{To evaluate possible selection effects, we test whether the observed $\xi_{\rm ion}$–overdensity trend could be driven by stellar mass, sSFR, or UV luminosity. 
Figure~\ref{fig:xion_M_sSFR}(a) shows a mild negative correlation between $\xi_{\rm ion}$ and stellar mass ($p = 0.14$), consistent with the trend reported in \citet{llerena_ionizing_2024}, noting that their shaded fit does not reflect the full scatter. 
Figure~\ref{fig:xion_M_sSFR}(b) reveals a positive correlation between $\xi_{\rm ion}$ and sSFR ($p = 0.03$), in agreement with previous work. 
However, as shown in Figures~\ref{fig:xion_M_sSFR}(c) and (d), we do not observe significant correlations between overdensity and either stellar mass ($p = 0.27$) or sSFR ($p = 0.82$) for galaxies \textit{in our sample}. 
In addition, we find no dependence of $M_{\rm UV}$ on overdensity (Figure~\ref{fig:MUV_density}). 
These results suggest that the observed $\xi_{\rm ion}$–overdensity trend is not likely driven by systematic variations in these physical properties across different environments.}




\end{document}